# Smartphone Sensors for Modeling Human-Computer Interaction: General Outlook and Research Datasets for User Authentication

Alejandro Acien, Aythami Morales, Ruben Vera-Rodriguez, Julian Fierrez
Biometrics and Data Pattern Analytics Lab (BiDA Lab), Universidad Autonoma de Madrid, Spain
{alejandro.acien, aythami.morales, ruben.vera, julian.fierrez}@uam.es

*Abstract*— **In this paper we list the sensors commonly available in modern smartphones and provide a general outlook of the different ways these sensors can be used for modeling the interaction between human and smartphones. We then provide a taxonomy of applications that can exploit the signals originated by these sensors in three different dimensions, depending on the main information content embedded in the signals exploited in the application: neuromotor skills, cognitive functions, and behaviors/routines. We then summarize a representative selection of existing research datasets in this area, with special focus on applications related to user authentication, including key features and a selection of the main research results obtained on them so far. Then, we perform the experimental work using the HuMIdb database (Human Mobile Interaction database), a novel multimodal mobile database that includes 14 mobile sensors captured from 600 participants. We evaluate a biometric authentication system based on simple linear touch gestures using a Siamese Neural Network architecture. Very promising results are achieved with accuracies up to 87% for person authentication based on a simple and fast touch gesture.**

## I. INTRODUCTION

In the last decade, smartphone devices have aroused a great interest in the scientific community due to the capacity of these devices to acquire, process, and storage a wide range of heterogeneous data. These data offer many possibilities and research lines, such as user authentication [1][2], health monitoring [3][4][5] or behavior monitoring [6][7][8][9] among others. Besides, the usage of mobile phones has spreaded out to the point that mobile lines exceeded world population in 2018 [10]. The main reason of this phenomenon is that mobile devices are absorbing many services that used to be consumed in other platforms (e.g. TV on demand, music/video streaming, social networks, e-books, videogames, business apps, etc) with also adding the possibility of usage anywhere/anytime.

In this paper we explore the potential of mobile devices to model human-machine interaction. The main contributions are: i) overview of signals and sensors employed in the literature to model human-machine interaction based on mobile interaction; ii) taxonomy of applications that can exploit these signals; iii) survey of research datasets used in the literature for smartphone user authentication; and iv) implementation of a new authentication system based on touch gestures using a Siamese Neural Network arquitecture with experiments on the recent HuMIdb[1] dataset. The proposed authentication system achieves accuracies up to 87% using simple linear touch gestures.

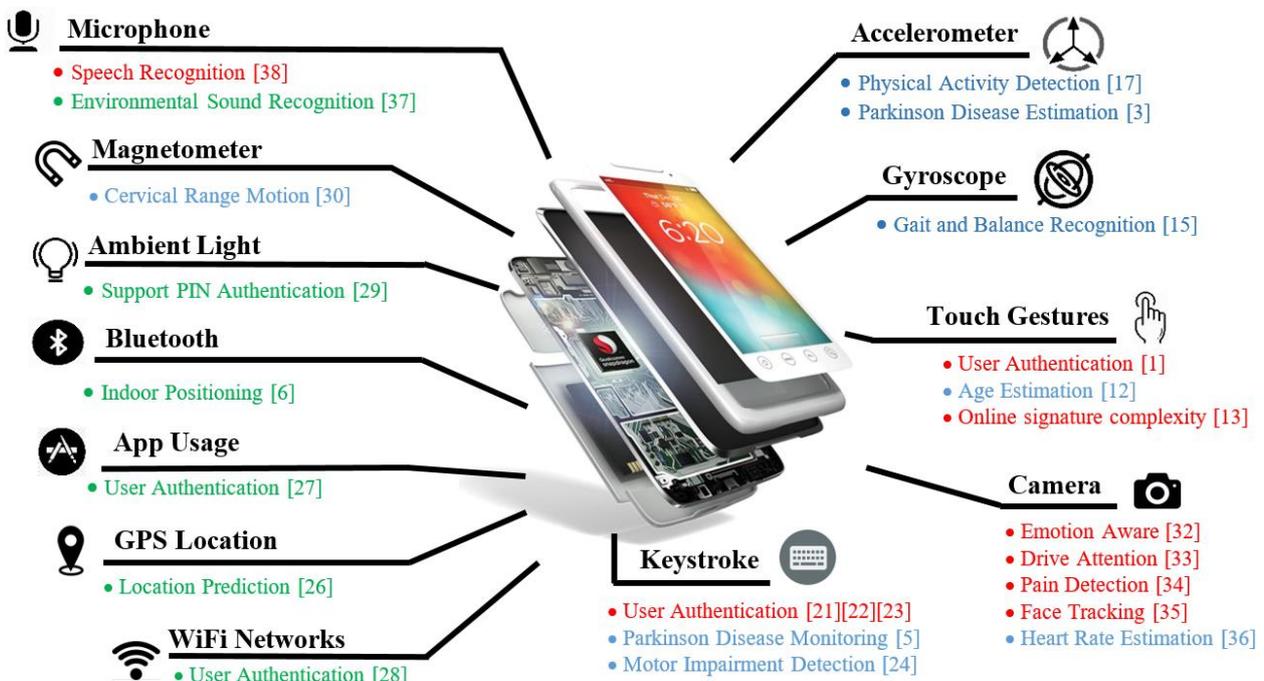

Fig. 1. A summary of the different sensors/signals of smartphone and example applications. In blue, applications that reveal neuromotor skills, in red, cognitive functions, and in green, applications revealing behaviors/routines.

---

[1] https://github.com/BiDAlab/HuMIdb

The remainder of the paper is organized as follows: Sec. II analyses the capacity of the different sensors of a smartphone to model human-machine interaction. Sect. III summarizes a representative selection of existing research datasets in this area. Sect. IV describes the HuMIdb database, first introduced in [11], which comprises 14 mobile sensors acquired from 600 users. Sect. V describes the proposed authentication method based on simple touchscreen interaction and the experimental results on HuMIdb. Finally, Sect. VI draws the final conclusions.

## II. MOBILE SENSORS FOR MODELING HUMAN-MACHINE INTERACTION

Smartphones contain many sensors such as accelerometer, gyroscope, gravity sensor, touchscreen, light sensor, WiFi, Bluetooth, camera, or microphone, among others, which can acquire information as the user is interacting with it or just carrying it. These sources of information can be used to model human-machine interaction and describe human features. Fig. 1 presents some examples of different research fields that exploit signals obtained or derived from mobile sensors.

- **Touch Gestures** involve all kinds of finger movements that we perform over the smartphone screen (e.g. swipe, tap, zoom). This biometric trait has already been used for user authentication [1]. More recently, the research community is focusing on the neuromotor patterns that can be extracted from touch gestures. As an example, Acien *et al.* [12] analysed the neuromotor patterns extracted from touch gestures to discriminate between children and adults, in order to adapt the content showed in the smartphone to the user age. In [13], the authors model the complexity of online signatures over smartphone touchscreens using the neuromotor patterns associated to touch gestures.

- **Accelerometer and gyroscope** are both useful to measure the movements that the smartphone is exposed to: the accelerometer measures the magnitude and direction of acceleration forces applied over the mobile device meanwhile the gyroscope measures orientation. Although these sensors have been studied for mobile user authentication with good results [14] in the last years, these qualities make both sensors traditionally useful for gait and balance recognition. For example, in [15] they employ these mobile sensors for user recognition trough simple gestures like answering a call in four different user states: standing, sitting, walking, and running. In another example, Gafurov *et al.* [16] extracted gait patterns from a mobile device attached to the lower part of the leg in three directions: vertical, forward-backward, and sideways motion. They achieved error rates between 5% and 9% for gait authentication combining all three acceleration measures. Accelerometer has been also studied to measure the daily physical activities with the main goal of changing people's sedentary lifestyle [17].

- **Keystroking** has been widely studied for user verification in physical keyboards by analyzing typing behavior, achieving great results with error rates under 5% [18][19]. On mobile device scenarios, the same concepts were adapted with little variations so far. As an example, in [20] a fixed-text keystroking system for mobile user authentication was studied using not only time and space based features (e.g. hold and flight times, jump angle, or drag distance) but also studying the hands postures during typing as discriminative information. In other works, Acien *et al.* [21][22] employed LSTM networks to take advantage of the temporal relationship between consecutive keys pressed during typing to develop a mobile keystroke authentication system. However, mobile keystroke systems present major limitations and their performance are far from the one achieved when using physical keyboards, unless they are combined with other mobile biometrics traits in multimodal mobile systems [22][23]. Very recently, taking advantage of the great acceptance and pervasive usage of smartphone devices, some works are studying mobile keystroking as a tool to remotely monitoring neuromotor impairment patients. In [5] an algorithm was presented to detect Parkinson's Disease by analyzing the typing activity on smartphones independently of the content of the typed text. This algorithm can help in clinical decision making by monitoring the patient motor status between hospital visits. In other work [24], the authors estimate motor impairments via touchscreen interactions during natural typing. The authors state that touchscreens can capture fine-movements of fingers during keystroking, an unsupervised activity of high frequency that can identify motor impairment.

- **WiFi, GPS and App Usage** are mobile signals that belong to behavioral-based profiling schemes due to their capacity to provide information about when and where we go and what we do. These mobile signals record the events (e.g. WiFi networks, Bluetooth signals, GPS locations, or application's name) and the timestamps of their occurrence. This discriminative information is considered as behavioral biometrics due to their capacity to detect variations in our daily routines [25]. As an example, Mahub *et al.* [26] developed a modified HMM to characterize mobile GPS location histories. They suggest that human mobility can be described as a Markovian Motion and they make predictions of the next user location taking into account the sparseness of the data and previous user locations. In a similar way, in [27] a variation of HMMs was studied to develop a user authentication mobile system by exploiting application usage data. The authors state that unforeseen events and unknown applications provide more discriminative information in the authentication process than the most common apps used. In [28] the authors perform a template-based matching algorithm for user authentication using the WiFi signals stored by the smartphone during the day. The fusion at score level with the accelerometer system achieve authentication error rates under 10%, showing the feasibility of WiFi signals to assist authentication on mobile devices.

- **Bluetooth** is a mobile signal similar to WiFi, which detects other Bluetooth beacons and the timestamps of occurrence. However, thanks to their low power consumption and the fact that works in a short range radio frequency, it is being studied for indoor positioning based on RSSI (Radio Signal Strength Indicator) Probability Distributions. For example, in [6] the authors applied the Weibull function to approximate the Bluetooth signal strength distribution in the data training phase.

- There are **Others Mobile Sensors** less obvious but also useful for modeling human machine interactions such as the light sensor, which measures the ambient-light level that the smartphone is exposed to. In [29] the authors demonstrate that minor tilts and turns in the smartphone cause variations of the ambient-light sensor information. These variations leak enough information to authenticate personal identification numbers. Another sensor, the magnetometer, has been also studied to measure the cervical range of motion on the horizontal plane using a smartphone placed on the head of the patient during a clinical trial [30].

| Ref. | Sensors | #Users | Sessions/user | Public | #Devices | Sensors: Performance |
|---|---|---|---|---|---|---|
| Fridman (2015) [40] | 4 (App, GPS, Sty, Web) | 200 | 5 Months | No | 200 | Stylometry: 5% EER |
| Mahub (2016) [41] *UMDAA-02* | 13 (Acc, Blu, Cam, GPS, Gyr, Key, Lig, Mag, Press, Prox, Temp, Touch, WiFi) | 48 | ~248 | Yes | 1 | Touch: 22% EER Cam (Face): 18% EER |
| Li (2018) [28] | 2 (Acc, WiFi) | 312 | 1 Month | No | 312 | Acc: 26% EER WiFi: 9% EER |
| Li (2018) [42] | 2 (Acc, Gyr) | 304 | ~90 Sessions | No | 304 | Acc + Gyr: 23% EER |
| Liu (2018) [43] | 5 (Acc, Gyr, Mag, Pow, Touch) | 10 | 3 Hours | No | 1 | Touch+Acc+Gyr+Mag: 5.5% EER |
| Shen (2018) [44] | 4 (Acc, Gyr, Mag, Ori) | 102 | 20 - 50 Days | Yes | 12 | Acc+Ori+Gyr+Mag: 4.5% EER |
| Deb (2019) [14] | 8 (Acc, GPS, Gra, Gyr, Key, LAc, Mag, Rot) | 37 | 15 Days | No | 37 | Key+GPS+Acc+Gyr+Mag+Acc+Gra+Rot: 99.98% TAR @ 0.1% FAR |
| Ramachandra (2019) [45] *SWAN* | 2 (Cam, Mic) | 150 | 6 Sessions | Yes | 1 | Cam+Mic: 24.56% ± 2.95 EER |
| Tolosana (2020) [46] *BioTouchPass2* | 3 (Acc, Gyr, Touch) | 217 | ≤ 6 Sessions | Yes | 217 | Touch: 6% EER |
| **Acien (2020) [11] *HuMIdb*** | **14 (Acc, Blu, GPS, Gra, Gyr, Key, LAc, Lig, Mag, Mic, Prox, Touch, Ori, WiFi)** | **600** | **≤ 5 Sessions** | **Yes** | **600** | **Touch: 13% EER** |

TABLE I. Relevant multimodal databases for research in human-machine interaction with special focus in mobile user authentication. Acc- Accelerometer, App- App usage, Blu- Bluetooth, Cam- Front camera, Gra- Gravity, Gyr- Gyroscope, Key- Keystroke, LAc- Linear Accelerometer, Lig- Light, Mag- Magnetometer, Mic- Microphone, Ori- Orientation, Pow- Power consumption, Press- Pressure, Prox- Proximity, Rot- Rotation, Sty- Stylometry, Temp- Temperature, Touch- Touch gestures, Web- Web browsing.

- Finally, the **Camera and Microphone** are two of the most important sensors of the mobile device. They take photos, selfies, record voice and sounds. These signals are being used for a wide range of research lines: user recognition [31], emotion aware [32], driver attention [33], pain detection [34], face tracking [35], heart rate estimation [36], environmental sound recognition [37], and speech recognition [38] among others. However, their capacity to collect private user information can be perceived as intrusive. As an example, the Spanish football company La Liga was fined for recording ambient sounds through clients smartphone to detect pirate streaming of their football channels without user's permission [39].

The literature demonstrates the potential of mobile sensors to model inner human features (e.g. cognitive functions, neuromotor skills, and human behaviors/routines). These devices become data hubs that can be used in many different applications related with the human-machine interaction.

### III. MOBILE DATASETS FOR MODELING HUMAN-MACHINE INTERACTION

In Table 1 we compare relevant multimodal databases for research in human-machine interaction with special focus in mobile user authentication. Friedman *et al.* [40] collected a multimodal mobile dataset that contains stylometry, app usage, web browsing, and GPS behavioral biometrics data from 200 subjects over a period of 5 months. They launched a mobile application that acquired the mobile data during natural mobile-human interactions (unsupervised scenario). They also mention that the main problem in the acquisition process was the battery drain, as the app was recording all the time the smartphone was on. UMDAA-02 [41] is a multimodal mobile database that includes 13 mobile sensors, see Table I. The data was collected during 2 months from 48 volunteers in an unsupervised scenario with 248 sessions per user in average using all of them the same smartphone (Nexus 5). In [28], WiFi and accelerometer mobile signals were collected from 312 participants in an unsupervised scenario during a period of 1 month. In [42], the same authors captured accelerometer and gyroscope mobile data from 304 participants in an unsupervised scenario with 90 sessions per user in average. In other work [43], the authors collect touch gestures, power consumption, accelerometer, gyroscope, and magnetometer mobile signals from 10 participants under laboratory conditions and with the same mobile device (supervised scenario) during a period of three hours. Shen *et al.* [44] collected accelerometer, gyroscope, orientation, and magnetometer mobile sensors from 102 subjects for active mobile authentication in an unsupervised scenario. During the acquisition they covered three smartphone-operating environments: hand-hold, table-hold, and hand-hold-walking. In a recent work [14], the authors collected up to 30 mobile sensor signals from 37 volunteers over a period of 15 days with an Android application that works in the background and acquire the mobile data passively. SWAN [45] is a multimodal mobile dataset that includes face, voice, and periocular human features extracted from 150 subjects during 6 sessions with an

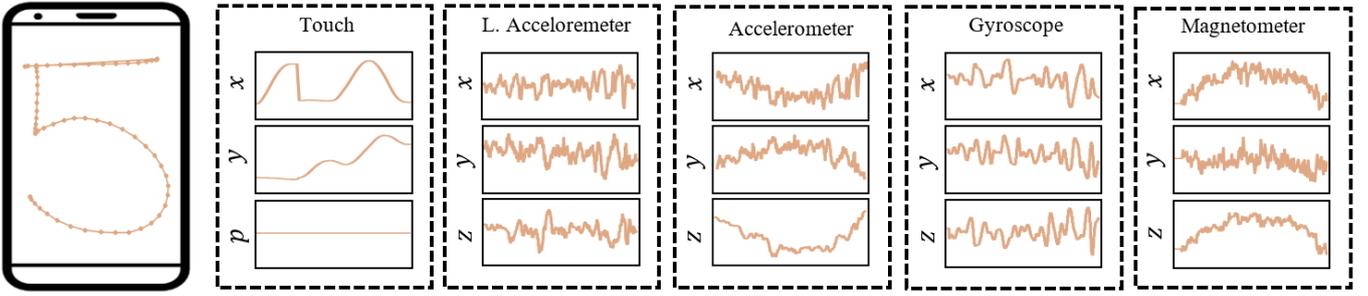

Fig. 2. Examples of time signals generated during a simple handwriting task (drawing a digit '5' with the finger).

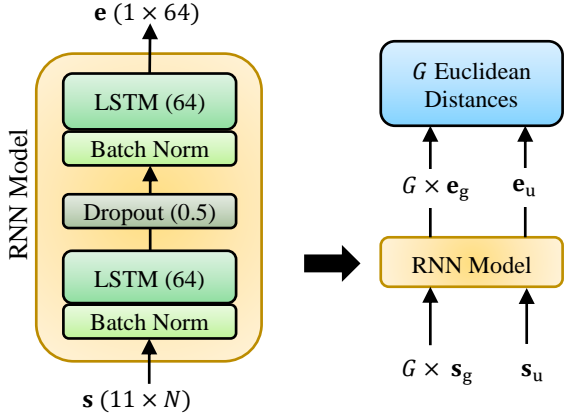

Fig. 3. The RNN architecture proposed for mobile touch authentication (left) and the scoring setup (right). The test scores are calculated as the average between $G$ scores obtained by comparing the embedding vectors ($\mathbf{e}_g$) of the gallery samples ($\mathbf{s}_g$) and the embedding vector ($\mathbf{e}_u$) of the unknown sample ($\mathbf{s}_u$).

iPhone 6S. The data collection was carried out in four different geographic locations: Norway, India, France, and Switzerland. Finally, in [46] the authors collected a database of mobile touch data named MobileTouchDB. The database is focused in mobile touch patterns and contains more than 64K character samples performed by 217 users during 6 acquisition sessions. They also acquired accelerometer and gyroscope signals under unsupervised conditions.

## IV. THE HuMI DATABASE: HuMIdb

In this section we describe the Human Mobile Interaction database: HuMIdb [2]. This database was presented in [11] and comprises more than 10 GB from a wide range of mobile sensors acquired under unsupervised scenario. The database includes 13 sensors during natural human-mobile interaction of 600 users (see Table I). For the acquisition, an Android application collected the sensor signals while the users complete 8 simple tasks with their own smartphones and without any supervision whatsoever (i.e., the users could be standing, sitting, walking, indoors, outdoors, at daytime or night, etc.) The different tasks were designed to reflect the most common interaction with mobile devices: keystrokes (name, surname, and a pre-defined sentence), taps (pressing a sequence of buttons), swipes (up and down directions), air movements (circle and cross gestures in the air), handwriting (digits from 0 to 9), and voice (record the sentence "I'm not a robot"). Additionally, there is a drag and drop button between tasks. See Fig. 2 for example signals. HuMIdb was designed for research in mobile biometrics and bot detection [11][47], among other areas related to human-machine interaction.

## V. USER AUTHENTICATION BASED ON TOUCH GESTURES

As an example application exploiting mobile sensors, in this section we explore a new authentication system based on the touch gestures acquired in HuMIdb. In particular, we employ for authentication the right-swipe gestures captured when the users scroll the drag and drop button to proceed between tasks. This is a common gesture used in many touch interfaces (e.g. unlock, next step confirmation).

### A. Neural Network Architeture

The architecture proposed to model swipe gestures is a RNN (Recurrent Neuronal Network) with a Siamese setup. We choose Siamese RNN models because they have proved to work well with temporal data [14][22]. The RNN architecture is composed by two LSTM (Long Short-Term Memory) layers of 64 units with batch normalization and dropout rate of 0.5 between layers to avoid overfitting (see Fig. 3 left for details). Additionally, each LSTM layer has a recurrent dropout rate of 0.2. The output of the model ($\mathbf{e}$) is an embedding vector of size 64 that contains discriminative information extracted from each swipe gesture to authenticate users. For this, we train the RNN model in a Siamese setup in which the model has two inputs (the two swipe samples to compare) and two embedding vectors as outputs. During the training phase, the RNN model learns to project embedding vectors from same user close to each other and embedding vectors from different users far from each other. Finally, to test our model we compute the Euclidean distance between pairs of embedding vectors ($\mathbf{e}_g$, $\mathbf{e}_u$); one from the genuine user that claims to authenticate in our system, called gallery sample ($\mathbf{s}_g$), and the unknown sample ($\mathbf{s}_u$) that we want to verify (see Fig. 3 right).

### B. Experimental Protocol

The interaction of the user with the touchscreen is defined by a time sequence $\{\mathbf{x}, \mathbf{y}, \mathbf{p}, \mathbf{t}\}$ with length $N$, composed by the coordinates $\{\mathbf{x}, \mathbf{y}\}$, the pressure $\mathbf{p}$, and the timestamp $\mathbf{t}$. The coordinates $\{\mathbf{x}, \mathbf{y}\}$ are normalized by the size of the screen. Then, we extract eleven temporal features adapted from [48][49] for on-line signatures: velocity, acceleration, jerk, and the Fourier transform for both axis $\{\mathbf{x}, \mathbf{y}\}$ plus the raw data $\{\mathbf{x}, \mathbf{y}, \mathbf{p}\}$. Note that we discard the timestamp $\mathbf{t}$ because it depends on the device and the network could be learning to discrimante among devices instead of users. Finally, the input of the RNN model ($\mathbf{s}$) is a feature set of size $11 \times N$ extracted for each swipe.

---

[2] https://github.com/BiDAlab/HuMIdb

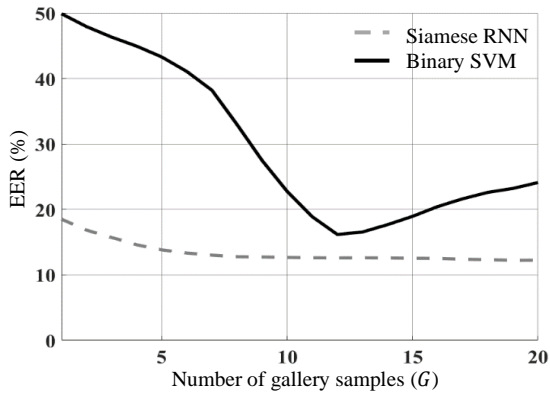

Fig. 4. Authentication based on touchscreen signals (single swipe): Error Rates (%) for increasing number of gallery samples (swipes) employed to model each user.

Training details: the best results were achieved with a learning rate of 0.05, Adam optimizer with $\beta_1 = 0.9$, $\beta_2 = 0.999$, $\varepsilon = 10^{-8}$; and the margin set to $\alpha = 1.5$ without learning decay. The model was trained after 30 epochs with 100 batches per epoch. Each batch has a size of 512 pairs. The pairs were chosen randomly but keeping the number of genuine and impostor pairs balanced in each batch. The model was built in *Keras-Tensorflow*. The RNN network is trained with 70% of HuMIdb users and tested with the remaining ones (open-set authentication paradigm). We want to highlight that there are a total of 30K swipe gestures in our experimental dataset. The size of the input features vector is set to $N = 100$, filling with zeros when the vectors are smaller and truncating in the opposite case.

Aditionally, we compare our proposed RNN model with our implementation of one of the best state-of-the-art systems traditionally employed in mobile touch authentication: global features extraction plus binary SVM (Support Vector Machines) classifiers with Gaussian kernel. In particular, we compute the global features presented in [50] (commonly used for online handwriting sequence modeling) and adapted for swipe biometrics in [1]. Mean velocity, max acceleration, distance between adjacent points, or angles are some examples of this subset of 29 features extracted. We employ the same experimental protocol for both systems. This means that we compute a binary classifier to authenticate each user by using the gallery samples as genuine training samples, and then we test with the remaining ones. This method allows us to compare the amount of user data each architecture needs.

*C. Results and Discussion*

The results are depicted in Fig. 4 in terms of EER, where EER (Equal Error Rate) refers to the value where False Acceptance Rate (percentage of impostors classified as genuine) and False Rejection Rate (percentage of genuine users classified as impostors) are equal. The curve shows the variation in the performance according to the number of gallery samples $G$ used to compute the score for each user, as the average of the Euclidean distances between the gallery samples and the unknown sample (see Fig. 3 right). For one-shot authentication ($G = 1$) our proposed system achieves an EER of 19%, and the performance improves when scaling up the number of gallery samples. In fact, with 6 gallery samples the EER is reduced to 13%, with no significant improvements for larger $G$. Comparing with the SVM architecture, we can observe that the Siamese RNN architecture obtains much better results.

These results prove the richness of touch gestures to model the interaction between humans and smartphones, in particular for user authentication. With a simple gesture (drag and drop) we have built an authentication system with good performance. Note that this performance is achieved under uncontrolled conditions including almost 600 different devices and non-supervised acquisition. Although these error rates can be considered high for some applications (e.g. in comparison with fingerprint or face authentication), the authentication based on touch gestures can be useful in continuous authentication scenarios where identity management is based on multiple evidences evaluated in a transparent setup [1][2][22][51].

## VI. CONCLUSIONS AND FUTURE WORK

We have explored the potential of mobile devices to model human-machine interaction. We presented a taxonomy of applications that can exploit the signals originated in those sensors in three different dimensions, depending on the main information content embedded in the signal or signals exploited in the application: neuromotor skills, cognitive functions, and behaviors/routines. We have overviewed the databases employed in the literature. These databases have been used traditionally for user authentication, but they provide signals useful for other applications as well beyond security and related to human behavior analysis. As example application, we experimented with HuMIdb, which to the best of our knowledge is the largest database of mobile sensor signals adquired during human mobile interaction to date, with 14 sensor signals collected from 600 users across 5 sessions and more than 600 devices involved. For that experiments we introduced a new method for user authentication based only on one touch gesture (drag and drop) and RNNs resulting in an error rate of 13%.

As future work we will increase the users in HuMIdb and improve our authentication system by merging it [51] with the information coming from other sensors like accelerometer, magnetometer, or gyroscope. Other applications beyond authentication like bot detection [47] and behavior tracking [52] will be also explored.

ACKNOWLEDGEMENTS

Support by projects: IDEA-FAST (IMI2-2018-15-two-stage-853981), PRIMA (ITN-2019-860315), TRESPASS-ETN (ITN-2019-860813), BIBECA (RTI2018 101248-B-I00 MINECO/FEDER), and BioGuard (Ayudas FBBVA 2017). Spanish Patent Application P202030066.